\documentclass[conference]{IEEEtran}
\IEEEoverridecommandlockouts
\usepackage{cite}
\usepackage{amsmath,amssymb,amsfonts}
\usepackage{algorithmic}
\usepackage{graphicx}
\usepackage{textcomp}
\usepackage{subcaption} % For subfigures
\usepackage{xcolor}
\usepackage{svg}
\def\BibTeX{{\rm B\kern-.05em{\sc i\kern-.025em b}\kern-.08em
    T\kern-.1667em\lower.7ex\hbox{E}\kern-.125emX}}
\begin{document}

\title{PrimeK-Net: Multi-scale Spectral Learning via Group Prime-Kernel Convolutional Neural Networks for Single Channel Speech Enhancement\\
% {\footnotesize \textsuperscript{*}Note: Sub-titles are not captured in Xplore and
% should not be used}
\thanks{*Corresponding author.}
}

\author{\IEEEauthorblockN{Zizhen Lin$^{1}$, Junyu Wang$^{2}$, Ruili Li$^{3}$\IEEEauthorrefmark{1}, Fei Shen$^{4}$, Xi Xuan$^{5}$}
\IEEEauthorblockA{$^{1}$School of Electronic Information, Sichuan University, Chengdu, China\\
Email: linzizhen17@163.com}
\IEEEauthorblockA{$^{2}$College of Intelligence and Computing, Tianjin University, Tianjin, China\\
Email: junyu\_wang21@tju.edu.cn}
\IEEEauthorblockA{$^{3}$School of Medicine, Tohoku University, Sendai, Japan\\
Email: li.ruili.t3@dc.tohoku.ac.jp}
\IEEEauthorblockA{$^{4}$School of Computer Science and Engineering, Nanjing University of Science and Technology, Nanjing, China\\
Email: feishen@njust.edu.cn}
\IEEEauthorblockA{$^{5}$Department of Linguistics and Translation, City University of Hong Kong, Hong Kong SAR, China\\
Email: xixuan3-c@my.cityu.edu.hk}
}

\maketitle

\begin{abstract}
Single-channel speech enhancement is a challenging ill-posed problem focused on estimating clean speech from degraded signals. Existing studies have demonstrated the competitive performance of combining convolutional neural networks (CNNs) with Transformers in speech enhancement tasks. However, existing frameworks have not sufficiently addressed computational efficiency and have overlooked the natural multi-scale distribution of the spectrum. Additionally, the potential of CNNs in speech enhancement has yet to be fully realized. To address these issues, this study proposes a Deep Separable Dilated Dense Block (DSDDB) and a Group Prime Kernel Feedforward Channel Attention (GPFCA) module. Specifically, the DSDDB introduces higher parameter and computational efficiency to the Encoder/Decoder of existing frameworks. The GPFCA module replaces the position of the Conformer, extracting deep temporal and frequency features of the spectrum with linear complexity. The GPFCA leverages the proposed Group Prime Kernel Feedforward Network (GPFN) to integrate multi-granularity long-range, medium-range, and short-range receptive fields, while utilizing the properties of prime numbers to avoid periodic overlap effects. Experimental results demonstrate that PrimeK-Net, proposed in this study, achieves state-of-the-art (SOTA) performance on the VoiceBank+Demand dataset, reaching a PESQ score of 3.61 with only 1.41M parameters.
\end{abstract}


\section{Introduction}

Sound significantly impacts the overall experience of audio systems, making speech enhancement algorithms essential for suppressing noise while restoring impaired speech. These algorithms improve video recording quality, voice communication clarity, and speech recognition accuracy \cite{pascual2017segan}.

Recent advancements in speech enhancement have been inspired by TSTNN \cite{tstnn}, often using the two-stage (TS) architecture. This includes structures like TS-Conformer \cite{cmgan_interspeech, wang23DPCFCS_interspeech, mpsenet_interspeech} and TS-Transformer \cite{DB-AIATinterspeech12, DPT-FSNetinterspeech29, stDPT}. Besides, These methods employ Dilated Dense Blocks (DDB) \cite{dilateddensenet} for encoding and decoding. The TS structure is widely used due to its efficiency to separate feature extraction for time and frequency dimensions. Notable examples include CMGAN \cite{cmgan_interspeech}, which demonstrates Conformer's effectiveness and introduces MetricGAN \cite{metricgan, metricgan+}, and MP-SEnet \cite{mpsenet_interspeech} which integrates CMGAN \cite{cmgan_interspeech} with PHASEN \cite{phasen} for phase spectrum estimation.

Although classic two-stage structures perform well in speech enhancement, their core components, DDB \cite{dilateddensenet} and Conformer \cite{conformer}, have limitations. Moreover, the potential of Convolutional Neural Networks (CNNs) \cite{CNNgradient} within this framework has not been fully explored. We identify several key issues with current state-of-the-art (SOTA) speech enhancement methods:

\begin{itemize}
    \item \textbf{Transformers \cite{vaswani2017attention} have $O(t^2)$ complexity.} Conformer \cite{conformer} utilizes Transformers' ability to capture long-range dependencies alongside convolutional modules for local feature extraction. However, Transformers' memory consumption scales with $O(t^2)$, which imposes many limitations on memory-constrained edge devices.
    
    \item \textbf{Previous methods have neglected the multi-scale characteristics of the spectrum.} The Conformer's \cite{conformer} Transformer and convolutional modules typically consider attention maps in fixed receptive fields, ignoring the multi-scale nature of the spectrogram.
\begin{figure*}[t]
  \centering
  \vspace{-0.4cm}
  \includegraphics[width=\linewidth]{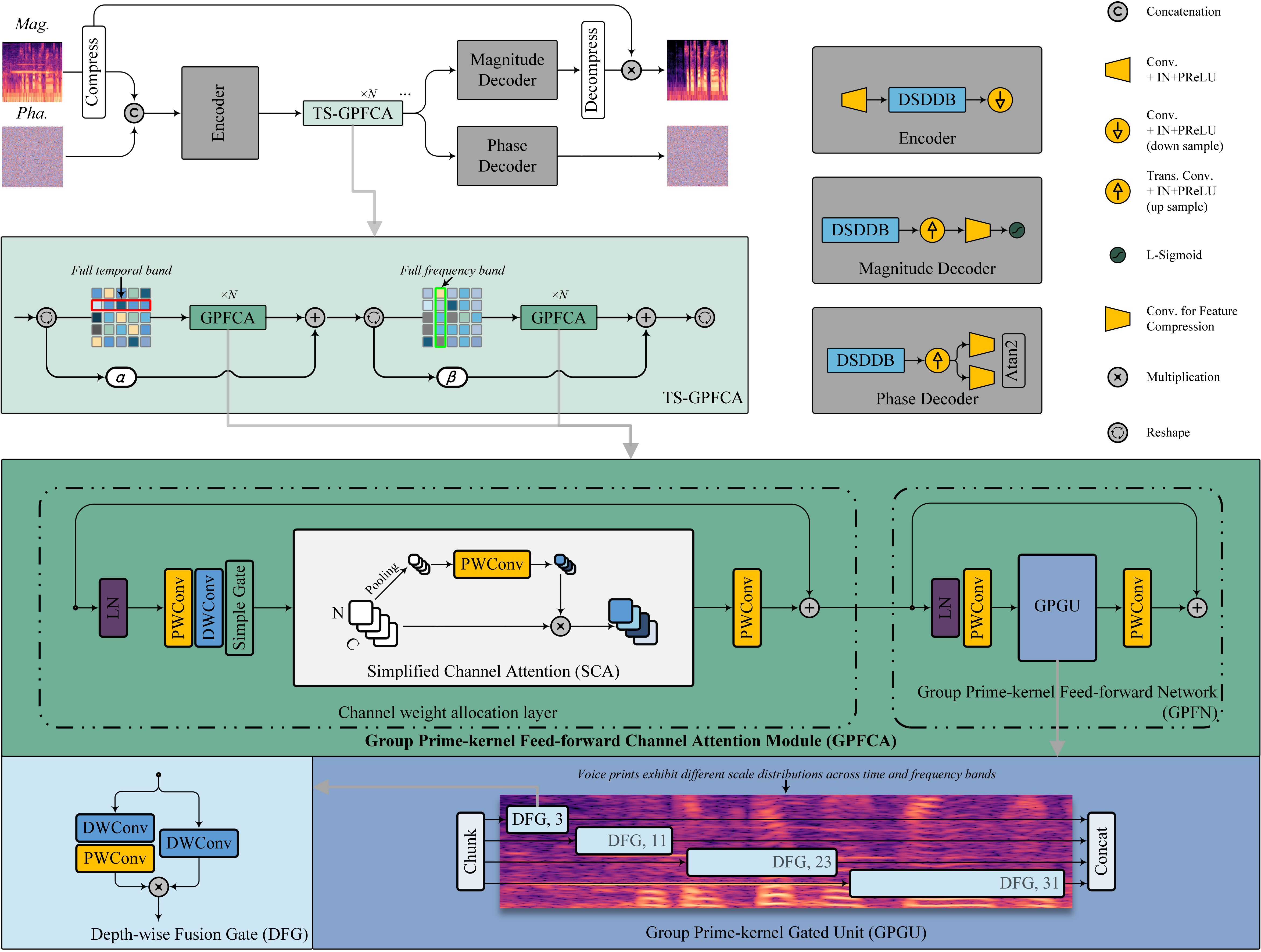}
  \caption{The overall architecture of the PrimeK-Net.}
  \label{fig:architecture}
  \vspace{-0.4cm}
\end{figure*}    
    \item \textbf{Efficiency issues with DDB \cite{dilateddensenet}. }Although DDBs play a crucial role in the encoder-decoder \cite{cmgan_interspeech,mpsenet_interspeech, wang23DPCFCS_interspeech, se-mamba, DB-AIATinterspeech12}, the channel dimension used in advanced methods increases with the number of layers while the initial $C$ is large, so we must consider the efficiency problem of $C^2$ in DDB.
\end{itemize}

To address these challenges, we developed PrimeK-Net, a new speech enhancement method based on efficient CNNs. It features the Group Prime-kernel Feed-Forward Channel Attention (GPFCA) module, which replaces Conformers in the Two-Stage structure. GPFCA uses prime number properties to avoid periodic overlap effects \cite{hifigan} and capture multi-scale time-frequency information while channel attention aggregating global information and facilitating channel interactions \cite{channelattensqueeze, NAFnetchen2022simple}. Compared to Conformer \cite{conformer}, GPFCA offers superior performance with fewer parameters and lower computational complexity, avoiding the \(O(t^2)\) complexity of Transformers \cite{vaswani2017attention}. Additionally, we introduce the Deep Separable Dilated Dense Block (DSDDB) as a streamlined modification of DDB \cite{dilateddensenet} for spectrum encoding and decoding.

Experiments performed on the VoiceBank + DEMAND \cite{voicebank} dataset demonstrate the effectiveness of our proposed method. Our PrimeK-Net\footnote{Source codes: https://github.com/huaidanquede/PrimeK-Net.} achieved a SOTA PESQ \cite{pesq} score of 3.61 with only 1.41M parameters, showcasing the potential of CNNs in speech enhancement.

\section{Method}
% \begin{figure*}[t]
%   \centering
%   \includegraphics[width=\linewidth]{arch编码器单独拆出SCA融入模块-覆盖式GPGU.png}
%   \caption{The overall architecture o the PrimeK-Net.}
%   \label{fig:architecture}
% \end{figure*}

This section covers the details of the Group Prime-kernel Feed-Forward Channel Attention (GPFCA) module, complexity and parameters of the Deep Separable Dilated Dense Block (DSDDB), and an overview of the employed loss function of our PrimeK-Net.

% The overall framework of PrimeK-Net shown in Figure \ref{fig:architecture} is similar to previous advanced methods \cite{cmgan_interspeech,mpsenet_interspeech,wang23DPCFCS_interspeech,se-mamba} Except for the internal details, which will not be discussed here.

\subsection{Group Prime-kernel Feed-forward Channel Attention Module (GPFCA)}

Inspired by NAFNet \cite{NAFnetchen2022simple}, we found that while retaining a module framework similar to that of the Transformer \cite{vaswani2017attention}, substituting the attention mechanism and feed-forward neural network (FFN) with more efficient structures can achieve competitive performance while maintaining linear complexity \cite{NAFnetchen2022simple}. To this end, we propose the Group Prime-kernel Feed-forward Channel Attention Module (GPFCA). As shown in Fig. \ref{fig:architecture} GPFCA combines a simple channel weight allocation layer with a Group Prime-kernel Feed-forward Network (GPFN) to enhance feature extraction and representation.

\textbf{Channel weight allocation layer.} As illustrated in the figure, the channel weight allocation layer of GPFCA is inspired by NAFBlock \cite{NAFnetchen2022simple}, and we have adapted it into a one-dimensional version suitable for two-stage structures. The Simplified Channel Attention (SCA) \cite{NAFnetchen2022simple} can be expressed as follows:
\vspace{-0.1cm}
\begin{equation}
\text{SCA}(\text{X}) = X \odot \text{PWC}(\text{AAP}(X))
\end{equation}

Adaptive average pooling (\(\text{AAP}\)) compresses each channel's sequence into a single value, while point-wise convolution (\(\text{PWC}\)) calculates channel weights, which are then multiplied with the original \(X\) for channel weight allocation, this scalar captures global information to some extent \cite{NAFnetchen2022simple,channelattensqueeze}, as it leverages the entire feature map of the current channel. Simplified Channel Attention avoids the \(O(t^2)\) complexity of Self Attention \cite{vaswani2017attention}, providing computational efficiency while maintaining key functions: aggregating global information and facilitating channel interactions \cite{channelattensqueeze, NAFnetchen2022simple}.

\textbf{Group Prime-kernel Feed-forward Network (GPFN).} We introduce GPFN for the feed-forward layer to enhance multi-scale feature extraction. GPFN employs a diverse group of Depth-wise Fusion Gates (DFGs) with prime-kernels, where "prime-kernel" refers to kernel sizes with prime numbers. The multi-scale design captures both transient and persistent speech variations at different time scales, addressing frequency features with varied scale distributions. The prime-kernel design reduces feature redundancy and improves accuracy by avoiding periodic responses, inspired by the Multi-Period Discriminator (MPD) in HIFI-GAN \cite{hifigan}.

Specifically, GPFN retains point-wise convolutions or linear layers for feature expansion and compression, replacing traditional gating with our proposed Group Prime-kernel Gated Unit (GPGU).

For GPGU, we chunk the output of the first point-wise convolution into four groups along the channel dimension:
% \vspace{-0.1cm}
\begin{equation}
{a}_{1}, {a}_{2}, {a}_{3}, {a}_{4} = \text{chunk}(X)
\end{equation}

Inspired by depth-wise gate in GDFN \cite{restormer}, we define the $i^{th}$ Depth-wise Fusion Gate (DFG) as:
% \vspace{-0.1cm}
\begin{equation}
\text{DFG}_{i}(X_{i}, k_{i}) = \text{PWC}(\text{DWC}_{k_{i}}(X_{i})) \odot \text{DWC}_{k_{i}}(X_{i})
\end{equation}

where \(\text{PWC}\) denotes point-wise convolution and \(\text{DWC}_{k_{i}}\) represents depth-wise convolution \cite{mobilenets} with kernel size $k_{i}$. For the grouped ${a}_{1}$, ${a}_{2}$, ${a}_{3}$, and ${a}_{4}$, DFG is computed with prime number kernel sizes of \{3, 11, 23, 31\} respectively. The outputs of the four DFGs are then concatenated along the channel dimension:

% \vspace{-0.3cm}
\begin{multline}
\text{GPGU}(X) = \text{CONCAT} ( \text{DFG}_{i} (a_{i}, k_{i}) \mid \\
a_{i} = \text{chunk}(X)[i], \; i \in \{1, 2, 3, 4\} )
\end{multline}

Finally, GPFN fuses the multi-scale features extracted by each DFG through point-wise convolution and adjusts the channel dimensions.

\subsection{Deep Separable Dilated Dense Block (DSDDB)}

\begin{figure}[t]
  \centering
  % \vspace{-0.4cm}
  \includegraphics[width=\linewidth]{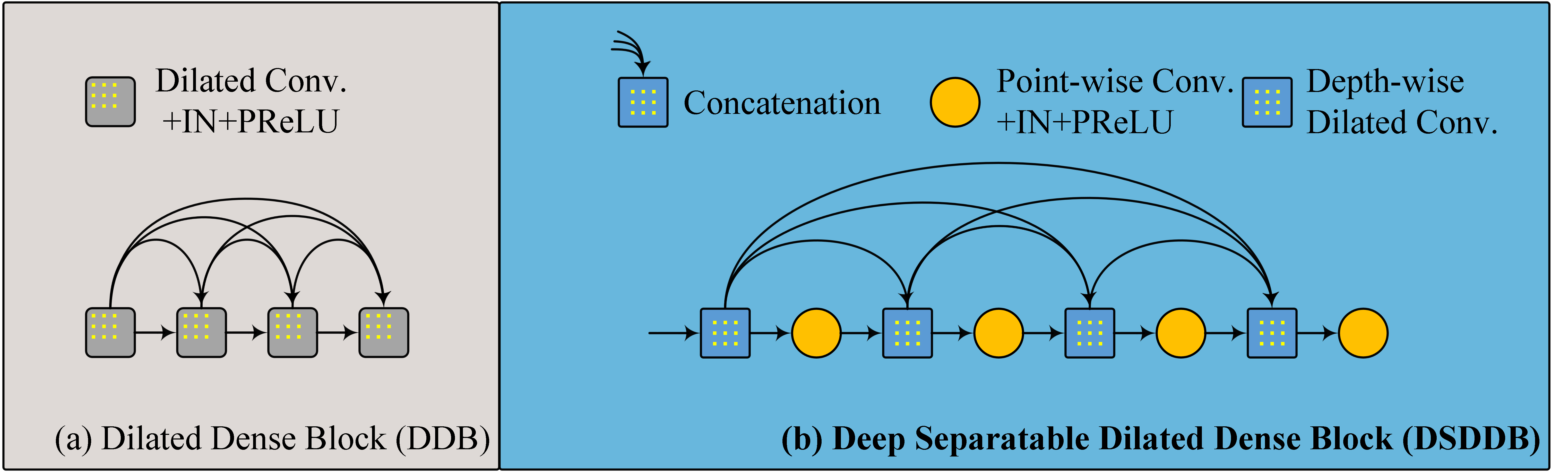}
  \caption{The internal details of the DDB (a) and DSDDB (b).}
  \label{fig:DSDDB}
  \vspace{-0.4cm}
\end{figure}

% Dense connection blocks based on convolutions were proposed in \cite{dilateddensenet}. These blocks utilize the concept of feature reuse, allowing the outputs of a given layer to be reused multiple times in subsequent layers. Direct connections between the given layer and subsequent layers help avoid the vanishing gradient problem in DNNs. 

Although the Dilated Dense Block (DDB) \cite{dilateddensenet}, as shown in Fig. \ref{fig:DSDDB} (a), performs exceptionally well in speech enhancement tasks, the number of parameters and computational complexity of the model increase significantly as the channel size of the model grows, particularly with an increase in the number of layers in the dilated dense convolution module.
 To address this issue, we propose the Deep Separable Dilated Dense Block (DSDDB), as shown in Fig. \ref{fig:DSDDB} (b).Similar to the strategy of depth-wise separable convolutions \cite{mobilenets}, each dilated convolution is decomposed into a combination of depthwise dilated convolutions and point convolutions. Considering only the convolution layers, the computational complexity for the $i$-th layer with input spectrum dimensions $t \times f$ for standard dilated convolution (DC) and depthwise separable dilated convolution (DSDC) can be expressed as:
% \vspace{-0.1cm}
\begin{equation}
\Omega(\text{DC}) = iC \times CK^2tf 
\end{equation}
\vspace{-0.1cm}
\begin{equation}
\Omega(\text{DSDC}) = iC \times K^2tf + iC \times Ctf
\end{equation}

where $C$ denotes the number of preset model channels, and $K$ is the size of the convolution kernel. For \( n \)-layer DDB and DSDDB, the computational complexity is:

\vspace{-0.1cm}
\begin{equation}
\Omega(\text{DDB}) = \sum_{i=1}^{n} iC^2K^2tf 
\end{equation}
\vspace{-0.1cm}
\begin{equation}
\Omega(\text{DSDDB}) = \sum_{i=1}^{n} iCK^2tf + \sum_{i=1}^{n} iC^2tf
\end{equation}

The parameter counts for DDB and DSDDB can be represented as:
\vspace{-0.1cm}
\begin{equation}
\text{P}(\text{DDB}) = \sum_{i=1}^{n} iC^2K^2
\end{equation}
\vspace{-0.1cm}
\begin{equation}
\text{P}(\text{DSDDB}) = \sum_{i=1}^{n} iCK^2 + \sum_{i=1}^{n} iC^2
\end{equation}

where $n = 4$, $C = 64$ and $K = 3$. Since $C^2$ is significantly larger than $K^2$, the complexity and parameter count of DSDDB are only \textbf{$12.7\%$} of DDB, with only a minor loss in performance.

\subsection{Loss Functions}

In this study, we employed two distinct loss functions to evaluate model performance. To facilitate comparison with previous state-of-the-art methods, we used the original loss function from MP-SENet \cite{mpsenet_interspeech}, denoted as \( L_{\text{old}} \). 

\begin{equation}
L _{ old }=\lambda_1 L _{\text {Metric. }}+\lambda_2 L _{\text {Mag. }}+\lambda_3 L _{\text {Pha. }}+\lambda_4 L _{\text {Com. }}
+\lambda_5 L _{\text {Time. }}
\end{equation}

Additionally, we applied an updated loss function from MP-SENet, referred to as \( L_{\text{new}} \). 

\begin{equation}
L _{ new }=\lambda_1 L _{\text {Metric. }}+\lambda_2 L _{\text {Mag. }}+\lambda_3 L _{\text {Pha. }}+\lambda_4 L _{\text {Com. }}
+\lambda_6 L _{\text {Con. }}
\end{equation}
This updated function is a minor modification of \( L_{\text{old}} \), which removes the time loss and incorporates a consistency loss \cite{scp}. This approach ensures a fair comparison with existing methods while also assessing the potential advantages of the updated loss function.

\section{Experiments}

\subsection{Datasets and Model setup}
In our study, we used the VoiceBank+DEMAND dataset \cite{voicebank}, which offers high-fidelity utterances. Speech was divided into two-second segments for the short-time Fourier transform, with an FFT size of 400, window length of 400, and hop length of 100. Training used a batch size of 2 with the AdamW \cite{kingma2014adam} optimizer for up to 1000k steps. Unless otherwise specified, the PrimeK-Net model defaults to using \textit{L$_{\text{new}}$}.

\subsection{Evaluation metrics}
We use widely accepted metrics to assess denoised speech quality: PESQ \cite{pesq} (scoring -0.5 to 4.5), MOS-based metrics (CSIG, CBAK, COVL) with a range of 1 to 5,. Higher scores indicate better performance.

\subsection{Result}

\begin{table}[htbp]
    \centering
    \caption{Comparison between MP-SENet, SE-Mamba, PrimeK-Net-S (1 GPFCA module) and PrimeK-Net (2 GPFCA modules). To ensure a fair comparison, the methods listed in this table \textbf{use the same loss function} (Without Consistency Loss and with Time Loss).}
    \label{tab:advance_net}
    \setlength{\tabcolsep}{8pt} % 设置列之间的间隙为4pt
    \begin{tabular}{lcccc}
    \hline
    Model & PESQ & CSIG & FLOPs & Para. \\
    \hline
    Noisy & 1.97 & 3.35 & - & -  \\
    \hline
    MP-SENet \cite{mpsenet_interspeech} & 3.50 & 4.73 & 74.52G & 2.05M \\
    SE-Mamba \cite{se-mamba}& 3.52 & 4.75 & 65.46G & 2.25M \\
    \textbf{PrimeK-Net-S \textit{L$_{\text{old}}$}} & 3.53 & 4.76 & \textbf{26.45G} & \textbf{0.79M} \\
    \textbf{PrimeK-Net \textit{L$_{\text{old}}$} } & \textbf{3.57} & \textbf{4.79} & 44.64G & 1.41M \\
    \hline
    \end{tabular}
\end{table}

To ensure a fair comparison of model performance, in Table \ref{tab:advance_net}, we utilize the same advanced framework and loss functions to evaluate MP-SENet, SE-Mamba, and PrimeK-Net. SE-Mamba and MP-SENet incorporate DDB in their encoder-decoder modules, while PrimeK-Net employs our DSDDB. All three models utilize a Two-Stage structure for deep feature extraction, with Conformer, Mamba, and GPFCA used internally in each model, respectively. Our experiments reveal that, under the same framework and loss function, PrimeK-Net not only significantly outperforms the other models but also achieves this with fewer parameters and reduced computational cost in terms of FLOPs.

% \vspace{-0.4cm}
\begin{figure}[ht]
    \centering
    \begin{subfigure}[b]{0.24\textwidth}
        \centering
        \includegraphics[width=\textwidth]{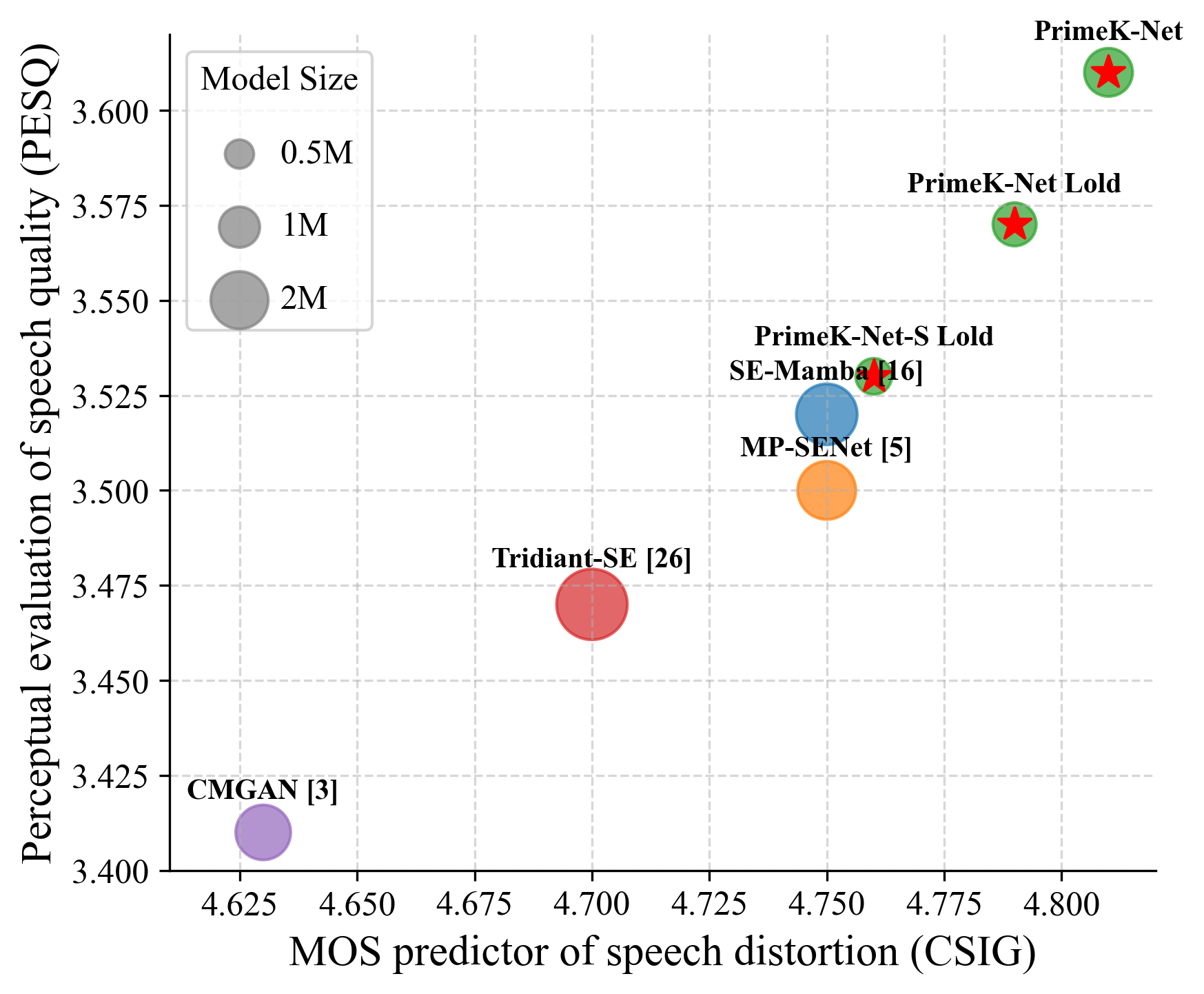}
        \caption{Comparison}
        \label{fig:image11}
    \end{subfigure}\hfill
    \begin{subfigure}[b]{0.24\textwidth}
        \centering
        \includegraphics[width=\textwidth]{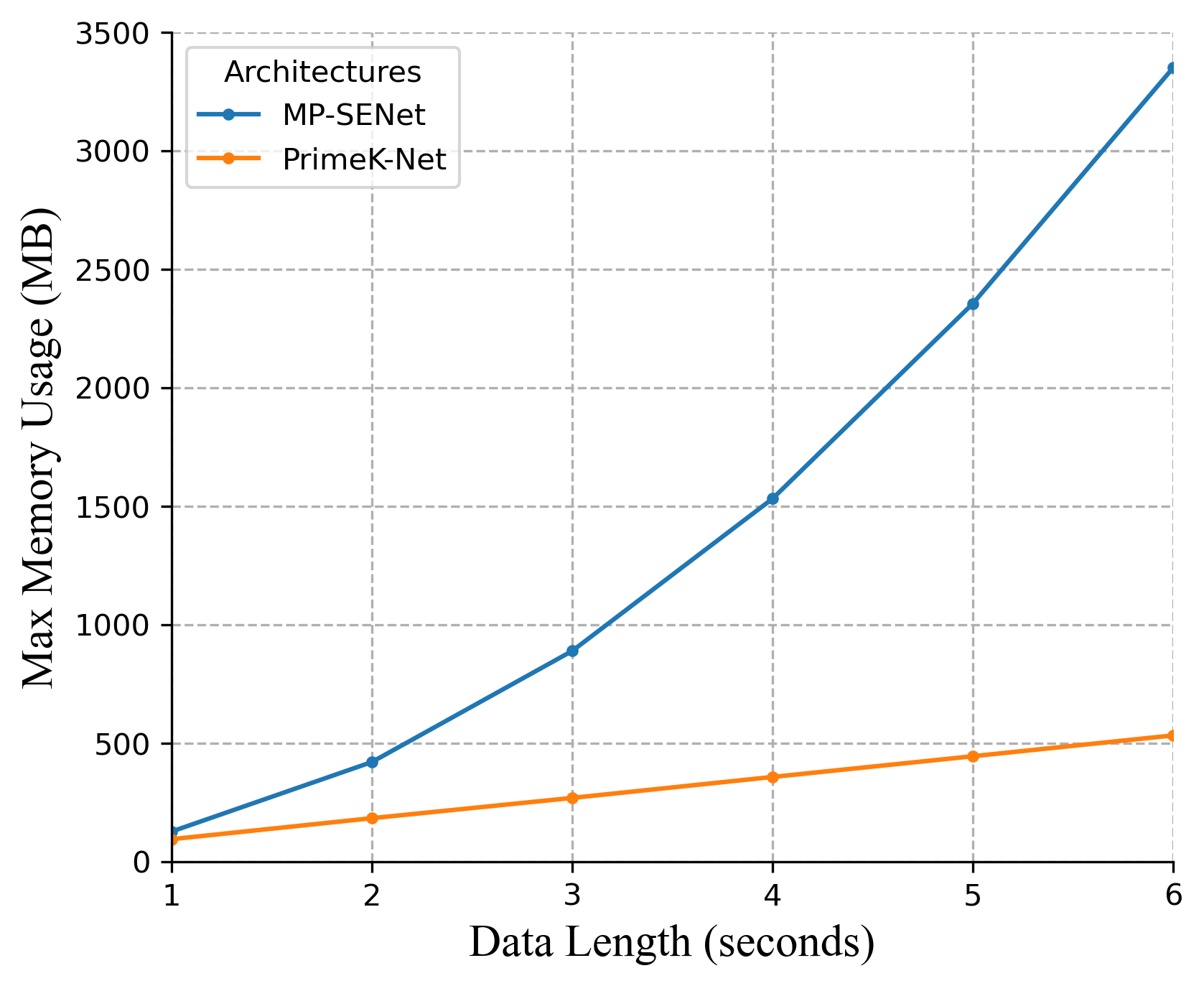}
        \caption{Memory Usage}
        \label{fig:image2}
    \end{subfigure}\hfill
    \caption{Comparison with other models.}
    \label{fig:comparemamery}
\end{figure}
% \vspace{-0.4cm}

In Fig. \ref{fig:comparemamery} (a), we compare several advanced models \cite{cmgan_interspeech, TridentSE_interspeech, mpsenet_interspeech, se-mamba}. In Figure~\ref{fig:comparemamery} (b), we illustrate the advantages of GPFCA in terms of linear complexity relative to MP-SENet \cite{mpsenet_interspeech}, which is based on Conformer \cite{conformer}.

\begin{table}[htbp]
    \centering
    \caption{Ablation Study Results on the VoiceBank + DEMAND Dataset}
    \label{tab:ablation}
    \setlength{\tabcolsep}{5pt} % 设置列之间的间隙为4pt
    \begin{tabular}{lccccc}
    \hline
    Model & PESQ & CSIG & CBAK & COVL & Para.\\
    \hline
    PrimeK-Net & \textbf{3.61} & 4.81 & \textbf{3.98} & 4.35  & 1.41M \\
    \hline
    GPFN$\rightarrow$Wide FFN & 3.51 & 4.74 & 3.93 & 4.27 &1.41M\\
    DSDDB$\rightarrow$DDB & \textbf{3.61} & \textbf{4.82} & \textbf{3.98} & \textbf{4.36} & 2.37M\\
    w/o SCA & 3.42 & 4.70 & 3.82 & 4.16 & \textbf{1.34M}\\
    w/o CL & 3.59 & 4.79 & 3.95 & 4.32 & 1.41M\\
    w/o CL + w/ TL (\textit{L$_{\text{old}}$}) & 3.57 & 4.79 & 3.97 & 4.30 & 1.41M\\
    
    \hline
    \end{tabular}
    
\end{table}

In Table \ref{tab:ablation}, we conduct ablation experiments and computational analysis on the module design of PrimeK-Net. Replacing our GPFN with a wide FFN from NafBlock \cite{NAFnetchen2022simple} shows that the performance improvement is not solely due to an increase in the number of parameters. Furthermore, we find that while DSDDB’s performance is slightly inferior to DDB, it significantly reduces both computational and parameter costs. Removing the SCA module subsequently leads to a decline in all metrics. Moreover, we investigated the impact of Consistency Loss (CL) and Time Loss (TL) on performance, where w/o CL + w/ TL is equivalent to \textit{L$_{\text{old}}$}.

\begin{table}[htbp]
    \centering
    \caption{Study of Depth-wise convolution kernel sizes in GPFN with single-scale, non-prime multi-scale, and prime multi-scale kernel sizes.}
    \label{tab:kernel_sizes}
    \setlength{\tabcolsep}{8pt} % Set column separation to 9pt
    \begin{tabular}{lcccc}
    \hline
    Kernel Size & PESQ & CSIG & FLOPs & Para. \\
    \hline
    (17, 17, 17, 17) & 3.57 & 4.77 & \textbf{44.64G} & \textbf{1.41M} \\
    (5, 15, 21, 27) & 3.59 & 4.79 & \textbf{44.64G} & \textbf{1.41M} \\
    (3, 11, 23, 31) & \textbf{3.61} & \textbf{4.81} & \textbf{44.64G} & \textbf{1.41M} \\
    \hline
    \end{tabular}
\end{table}

In Table \ref{tab:kernel_sizes}, we perform ablation experiments on the kernel size selection for DFG within PrimeK-Net. We compare the performance differences among single-scale, non-prime multi-scale, and prime multi-scale configurations with equivalent computational and parameter costs.

\section{Conclusions}

In this study, we propose PrimeK-Net, which utilizes CNNs for advanced speech enhancement. Our GPFCA module combines global, channel, and multi-scale time-frequency information while using prime numbers to avoid periodic overlap. We also introduce DSDDB to reduce the computational cost of Dilated Dense Blocks. Experiments on the VoiceBank+Demand dataset show that our model is more lightweight and outperforms state-of-the-art models. In the future, We will apply our approach to other tasks such as ASR and speech separation.

\bibliographystyle{ieeetr} %ieeetr为一种IEEE期刊标准的参考引用格式，其余还有多种选择，可根据自己所投期刊进行修改
\bibliography{mybib}%%reference就是你所命名的bib文件的文件名字

% \begin{thebibliography}{00}

% \bibitem{b1} Wang, Kai, Bengbeng He, and Wei-Ping Zhu. "TSTNN: Two-stage transformer based neural network for speech enhancement in the time domain." ICASSP 2021-2021 IEEE International Conference on Acoustics, Speech and Signal Processing (ICASSP). IEEE, 2021.
% \bibitem{b2}
% \end{thebibliography}
% \vspace{12pt}
% \color{red}
% IEEE conference templates contain guidance text for composing and formatting conference papers. Please ensure that all template text is removed from your conference paper prior to submission to the conference. Failure to remove the template text from your paper may result in your paper not being published.

\end{document}